%% file: main.tex
\documentclass[%
reprint,
nofootinbib,
amsmath,amssymb,
aps,
floatfix,
]{revtex4-1}
\pdfoutput=1
\usepackage{graphicx}% Include figure files
\usepackage{dcolumn}% Align table columns on decimal point
\usepackage{bm}% bold math
\usepackage{csquotes}
\begin{document}

% \preprint{APS/123-QED}

\title{Electronic Conduction Mechanisms in GeSe$_3$:Ag and Al$_2$O$_3$:Cu.}

\author{K. N. Subedi}
\author{Kiran Prasai}
\author{D.A. Drabold}
\email{drabold@ohio.edu}
\affiliation{
 Department of Physics and Astronomy, Nanoscale and Quantum Phenomena Institute, Ohio University, Athens, OH 45701, U.S.A.\\
}
\date{\today}% It is always \today, today,
             %  but any date may be explicitly specified

\begin{abstract}
In this paper, we discuss the atomistic structure of two conducting bridge computer memory materials, including Cu-doped alumina and silver-doped GeSe$_3$. We show that the Ag is rather uniformly distributed through the chalcogenide glass, whereas the Cu strongly clusters in the alumina material. The copper oxide system conducts via extended state conduction through Cu atoms once the concentration becomes high enough to form connected Cu channels. What is more, the addition of Cu leads to extended states throughout the large alumina (host) optical gap. By contrast, Ag in the selenide host is not strongly conducting even if one imposes a very narrow nanowire geometry. All of these results are discussed using novel techniques for computing the conduction active parts of the network.
\end{abstract}
\maketitle

\section{\label{sec:level1}Introduction}

Chalcogenide materials (elemental or alloyed systems involving S, Se and Te) are among the most interesting in science. They appear to offer an inexhaustible range of applications and scientific puzzles that are both challenging and interesting \cite{zakery_optical_prop_chalco_review, burno_chalco_review}. For these reasons, there is a rich literature both on experimental research, and theory, which in recent years mostly means computer simulations. Here, we give a few representative citations to the literature; though this paper is by no means a thorough survey of a vast and important field. We essentially limit ourselves here to a discussion of theory and modeling. As in most scientific fields, it is experimental research that is the primary driver of understanding and innovation, see for example Ref. \cite{adler_threshold_switching_chalco}.\\
For completeness, we note a few of the applications of chalcogenides. GeSbTe materials in appropriate combination are the basis of Phase Change Memory (PCM) devices \cite{wong_pcm,wuttig_pcm}, now in practical use for low-power applications like cell phones. The basic point is that the amorphous and crystalline phases exhibit significantly different electronic conductivities, and this contrast is the basis of the memory functionality. Impressive calculations have shown that it is possible to directly simulate the physical processes of amorphization and crystallization with accurate (density functional) methods \cite{hegedus_pcm_GeSbTe,jones_pcm_concepts_misconceptions}, such devices are also relatively radiation hard \cite{koinstantinou_radiation_tolerence_Ge2Sb2Te5}. Entirely new avenues are opening up on subjects like neuromorphic computing \cite{nandakumar_pcm_neuromorphic_computing}, and superlattice materials\cite{Li_pcm_superlattice}, among others. Surprisingly, chalcogenides are also the basis of another altogether different approach to non-volatile computer memory based on the fast-ion conducting properties \cite{angell_fast_ion_transport} of materials like GeSe3:Ag (but one example of many of a so-called conducting bridge (CBRAM) materials \cite{waser_nanoionics,valov_electrochemical_metallization}).
Here, the Ag ions are remarkably mobile, and the dynamics are directly accessible from computer simulation \cite{tafen_chalco,choudary_Ag_dynamics}. Electrochemical cells allow the controllable and ultrafast formation and destruction of electronically conducting paths (built typically from transition metals such as Ag or Cu) in these cells, which provides a stark contrast in electrical conductivity, and thus another attractive non-volatile memory device. Elliott pointed out a remarkable \enquote{optomechanical} effect in amorphous GeSe \cite{optomechanical_stuchlik}. Boolchand found an interesting and still poorly understood \enquote{Intermediate Phase}, wherein finite composition windows exhibit fascinating thermal,and other behavior \cite{boolchand_chalcogen_i,boolchand_chalcogen_ii}. Photo-response and manipulation \cite{stone_photoresponse} of chalcogenides is another rich area, and the phenomenon of \enquote{photomelting} is an example of an electronic/optical destabilization of a glassy network not due to simple heating \cite{hisakuni_optical_microfabrication,tanaka_photoinduced_process_chalcogen,tanaka_photoinduced_phenomenon,kovolov_loss_of_longerange_order}. This is a highly incomplete list.\\
Chalcogenides are realistically modeled using {\it ab initio} methods, or schemes such as accurate Machine Learning  potentials \cite{volker_ML_potentials,gabardi_2018,mocanu_ML_interatomic_potential} (based upon, and essentially converged to \textit{ab initio} interactions). For questions of atomistic processes, empirical potentials are not up to the task: the chemistry and energetics are too delicate. Frankly, care is needed in interpreting the literature when physical claims are drawn from unphysical potentials. Unfortunately, accurate calculations, even those based upon Machine Learning, are not computationally cheap. The good news, however, is that the simplest scheme of \enquote{melt-quenching} (modeling a glass by simulating a liquid and then simulating  quench) usually works well, at least for the general features and topology of the materials, and as mentioned above, even allows for the direct exploration of phase transitions \cite{dadepj}. We should also mention that to \enquote{really get the topology right}, plain vanilla DFT is not suitable, and careful workers such as Massobrio \cite{massobrio,evelyne}  and Micolaout \cite{micoulaut} have shown the importance even of van der Waals corrections. It is sometimes helpful to use modeling schemes that employ experimental information as part of the model determination process. In spirit, this is like \enquote{Reverse Monte Carlo}, which is much more powerful when merged with \textit{ab initio} interactions in an unbiased way “Force Enhanced Atomic Refinement” (FEAR) \cite{anup_fear}.

The two examples of this paper will involve CBRAM technology. Such materials consist of an oxidizable electrode {\it eg.} Ag or Cu and an inert counter electrode eg. Pt, W or TiN. The switching mechanism of these devices is the formation/dissolution of a conducting filament under external bias \cite{kozicki_2016,mueller}. The nanoscale filament consists of metal ions, as experimentally confirmed by Transmission Electron Microscopy (TEM) and Energy-Dispersive X-ray Spectroscopy \cite{fujii_cf_exp,liu_cf_exp}. There are many material realizations exhibiting attractive switching and stability for these devices ranging from chalcogenides \cite{chalco1,chalco2_gese2,gese}, oxides \cite{sio2_cbram,ta2o5_oxidei,ta2o5_oxideii,al2o31,al2o32} to bilayer materials \cite{bilayer1,bilayer2}. In this chapter, we study two materials: GeSe$_3$:Ag and Al$_2$O$_3$:Cu. We perform \textit{ab-initio} calculations of these materials with varying concentrations of transition metals (Ag or Cu) and study electronic conduction mechanisms. We exploit the Kubo-Greenwood formula to compute a spatially projected conductivity (SPC) and visualize the conduction-active networks in these materials.
\section{\label{sec:level2} Methods}
\subsection{\label{sec:level2_1}\textit{Ab-initio} simulations}
The models presented in this work are obtained by performing \textit{ab-initio} molecular dynamics (AIMD) simulations using the Vienna Ab-initio Simulation Package (VASP) \cite{vasp}. We follow the quench-from-melt scheme \cite{dadepj} where an initial configuration is heated to a high temperature above its melting point and cooled to room temperature. For all models, projector-augmented wave (PAW) \cite{paw1,paw2} potentials were used. Periodic boundary conditions were applied, and the temperature was controlled by a Nos\'e-Hoover thermostat \cite{nose_thermostat}.

\begin{figure}[!h]
 \begin{center}
 \includegraphics[width=1.7in]{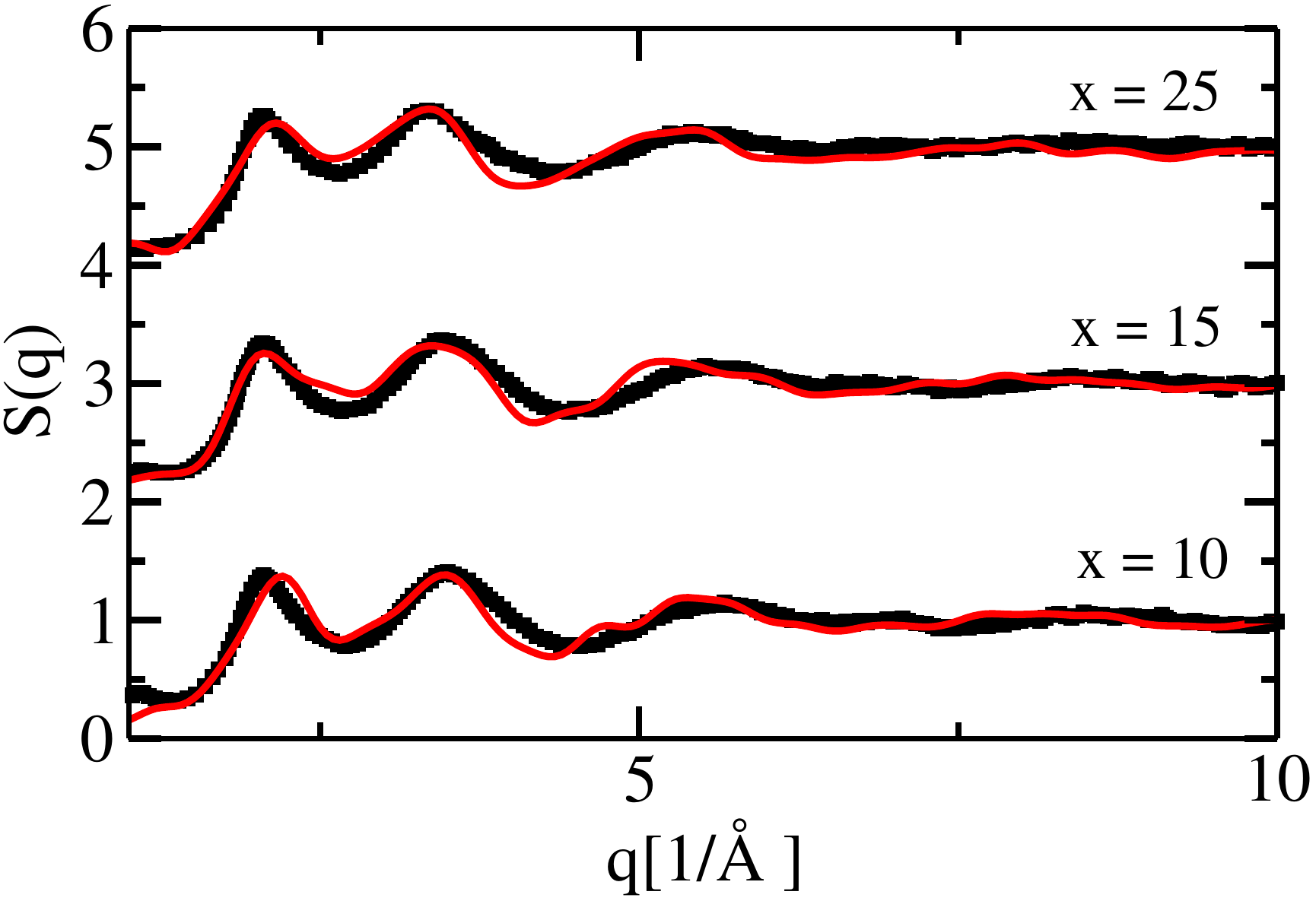}
 \includegraphics[width=1.4in]{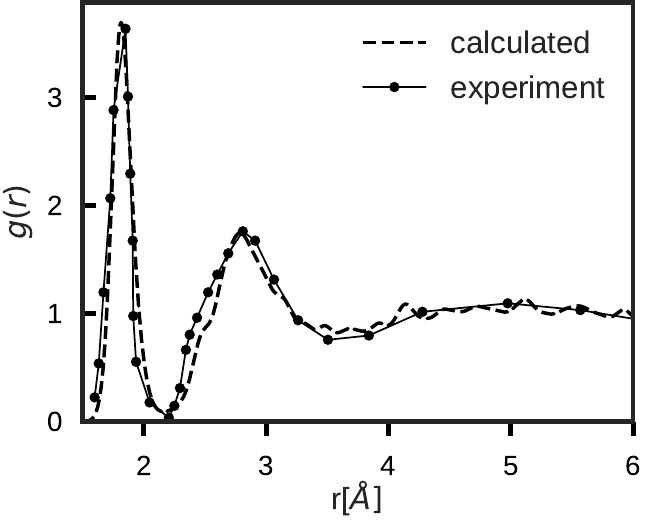}
 \caption{ LEFT: Structure factors ($S(q)$) of (GeSe$_3$)$_{100-x}$Ag$_x$ models (solid red line) compared with experiment (black squares) \cite{piarristeguy2000}. For the sake of clarity, vertical axis is shifted by 2 for x=15 and by 4 for x=25. RIGHT: Radial distribution function ($g(r)$) of alumina (dashed line) compared with experiment (solid circled line) \cite{LAMPARTER1997405}.}
 \label{sq_gr_compare_exp}
 \end{center}
 \end{figure}

We prepared models of (GeSe$_3$)$_{1-n}$Ag$_n$ for values of $n$ = 0.1, 0.15, 0.25 and 0.35. For compositions with $n$ = 0.1, 0.15 and 0.25, we employed a supercell containing 240 atoms at experimental densities \cite{piarristeguy2000,mirandou2003}. For Ag concentration of 35\% ($n=$0.35), we guessed the density, and later did a zero-pressure minimization to correct the guess. Plane waves of up to 350 eV were used to expand the Kohn-Sham orbitals, and PAW potentials with gradient corrections \cite{perdew1996} were used. The starting configurations for AIMD were obtained by performing a separate quench-from-melt simulation using the classical two-body potential of Iyetomi {\it et al} \cite{vasistha}. Using AIMD, these configurations were equilibrated at 2000 K for 4 ps, cooled to 1200 K over 12 ps, equilibrated at 1200 K for 6 ps, and then cooled to 300 K over 18 ps. All systems were subsequently equilibrated at 300 K for 10 ps. Finally, the models were relaxed to their nearest energy minima using a conjugate gradient algorithm \cite{numerical_recipe}. The structure factors of the resulting models are plotted with comparisons to experimental structure factors in left plot of figure \ref{sq_gr_compare_exp}.
\begin{table}[!h]
\begin{center}
\caption{\bf System Stoichiometries and Densities}\label{tab1}
\begin{tabular*}{0.9\linewidth}{@{\extracolsep\fill}cccccc@{\extracolsep\fill}}
\hline
\textbf{x}&\textbf{Mol. Formula}&\textbf{N$_{Ge}$,N$_{Se}$,N$_{Ag}$}&\textbf{$\rho$(gm/cm$^{3}$)}\\
\hline
10 & (GeSe$_{3}$)$_{0.90}$Ag$_{0.10}$ & 54, 162, 24 & 4.98  \\
15 & (GeSe$_{3}$)$_{0.85}$Ag$_{0.15}$ & 51, 153, 36 & 5.03  \\
25 & (GeSe$_{3}$)$_{0.75}$Ag$_{0.25}$ & 45, 135, 60 & 5.31  \\
35 & (GeSe$_{3}$)$_{0.65}$Ag$_{0.35}$ & 39, 117, 84 & 5.52  \\\hline
\textbf{x}&\textbf{Mol. Formula}&\textbf{N$_{Al}$, N$_{O}$, N$_{Cu}$}&\textbf{$\rho$(gm/cm$^{3}$)}\\
\hline
10 & (Al$_2$O$_{3}$)$_{0.90}$Cu$_{0.10}$ & 72, 108, 20 & 3.75  \\
20 & (Al$_2$O$_{3}$)$_{0.80}$Cu$_{0.20}$ & 64, 96, 40 & 3.99  \\
30 & (Al$_2$O$_{3}$)$_{0.70}$Cu$_{0.30}$ & 56, 84, 60 & 4.82  \\
\hline
\end{tabular*}
\end{center}
\end{table}
For Cu-doped alumina, we generated four amorphous models of composition (Al$_2$O$_3$)$_{1-n}$Cu$_{n}$ where $n$ is 0.0, 0.1, 0.2 and 0.3. We modeled $a$-Al$_2$O$_3$ at a density of 3.175g/cm$^3$ as described in other literature \cite{gutierrez,vasistha_al2o3}. For the Cu-doped models, we referred to the literature \cite{cudoped_reference} to make an initial guess of density and later on performed zero-pressure relaxations to obtain the final density. Plane waves up to 420 eV were used and PAW potentials with local density approximation \cite{LDA} were used. Each model was annealed at 3500 K for 7.5 ps, followed by cooling to 2600 K at a rate of 0.27 K/fs \cite{cooling_rate_al2o3} and an equilibration for 10 ps at 2600 K. Each model was then quenched to 300 K at 0.27 K/fs and equilibrated for 10 ps at 300 K \footnote{Of course, this heating/cooling/annealing schedule is somewhat arbitrary, but this choice seems to offer a reasonable balance between experimental realism and computational demand.}. The models were relaxed using the conjugate gradient algorithm. Zero-pressure relaxations were used to determine proper densities for the Cu-doped models. The radial distribution function $g(r)$ is compared with experiment as shown in right plot of Fig. \ref{sq_gr_compare_exp}. The final densities for all the models are given in table \ref{tab1}\\
\subsection{\label{sec:level2_2} Spatial Projection of Electronic Conductivity}
Here, we discuss computing the spatial projection of the electronic conductivity based on a novel spatial decomposition of the Kubo-Greenwood formula (KGF) \cite{kubo,greenwood,martin_2004}. This approach can estimate what parts of a complex material are active in electronic conduction (or absorb light at finite frequency). We discuss this method in detail in a recent paper \cite{rrl_prasai18}. The KGF has been exploited in tight-binding and DFT computations of conductivity in liquids \cite{liquid_silicon} and solids \cite{carbon_KGF}. The idea is to spatially decompose the conductivity in terms of Kohn-Sham orbitals in real space. We describe two ways to achieve this, one directly from the KGF and the other in terms of the eigenvectors of a Hermitian matrix described below.\\
The diagonal elements of the conductivity tensor for a specific k-point and frequency can be written in a compact form as equation (\ref{kgf_eqn}) (also discussed in \cite{simplified_kgf}).
\begin{equation}
\begin{aligned}
\sigma_{k}(\omega) =\frac{2\pi e^2}{3m^2\omega \Omega}\sum_{i,j} \sum_\alpha [f(\epsilon_{i,k})-f(\epsilon_{j,k})]  \\
{\mid \langle \psi_{j, k}|p^\alpha|\psi_{i,k} \rangle \mid}^2 \delta(\epsilon_{j,{ k}}-\epsilon_{i,{k}}-\hbar \omega)
\label{kgf_eqn}
\end{aligned}
\end{equation}
In the above Eq. (\ref{kgf_eqn}), \textit{e} and \textit{m} represent the charge and mass of electron respectively. $\Omega$ represents the volume of the supercell chosen for the simulation. For the rest of the discussion, we drop the vector notation for space points ($x$) or special k-points ($k$). We average over the diagonal elements of conductivity tensor. $\psi_{i,k}$ indexes the Kohn-Sham orbitals associated with energies $\epsilon_{i, k}$ and $f(\epsilon_{i,k})$ denotes the Fermi weight. $p^\alpha$ is the momentum operator along $\alpha$ direction. Let $$g_{ij}({ k},\omega)=\frac{2\pi e^2}{3m^2\omega \Omega}\sum_{i,j} [f(\epsilon_{i,k})-f(\epsilon_{j, k})]\delta(\epsilon_{j,{ k}}-\epsilon_{i,{ k}}-\hbar \omega).$$ Then, suppressing the explicit dependence of $\sigma$ on $k$ and $\omega$, the conductivity can be expressed as double spatial integrals:
\begin{small}
\begin{equation}
\sigma=\sum_{i,j,\alpha} g_{ij}\int d^3x\int d^3x^\prime[\psi_{j}^*(x)p^\alpha\psi_i(x)][\psi_{i}^*(x^\prime)p^\alpha\psi_j(x^\prime)]
\label{int_sigma}
\end{equation}
\end{small}
If we define $\xi_{ij}^\alpha(x)=\psi_i^*(x)p^\alpha\psi_j(x)$ as complex-valued functions 
on real space grid points (say $x$) with equal spacing of width \textit{h} in three dimensions, then we can approximate the integrals as a sum on grids (taken for simplicity here as uniform within a cube, with spacing $h$). Thus, Eq. (\ref{int_sigma}) can be written as 
\begin{equation}
\sigma \approx h^6\sum_{x,x^\prime}\sum_{i,j,\alpha} g_{ij}\xi_{ji}^\alpha(x)\xi_{ij}^\alpha(x^\prime)
\label{sigma_approx}
\end{equation}
If we define a Hermitian, positive semi-definite matrix 
\begin{equation}
\Gamma(x,x^\prime) = \sum_{i,j,\alpha} g_{ij}\xi_{ji}^\alpha(x)\xi_{ij}^\alpha(x^\prime)
\label{gamma}
\end{equation}
then, it follows from Eq. (\ref{sigma_approx}) that $\sigma=\sum_{x,x^\prime}\Gamma(x,x^\prime)$ as $h \rightarrow 0$. 

Equation (\ref{gamma}) suggests a way to spatially decompose the $\sigma$ at each grid point as 
\begin{equation}
\zeta(x) = \mid\sum_{x^\prime}\Gamma(x,x^\prime)\mid.
\label{zeta_x}
\end{equation}

  $\Gamma(x,x^\prime)$ decays because of destructive quantum mechanical wave interference, very much in the spirit of the decay of the single-particle density matrix, which is a basis of linear scaling methods in the theory of materials. This logic was clearly expressed by Kohn with his \enquote{principle of nearsightedness} \cite{kohn_Order_N_method,xiaodong} developed for the density matrix. Practically the same reasoning applies here, except that a different energy space is summed (one within $kT$ of the Fermi level in our work, as opposed to a sum over all states up to the Fermi level or Fermi surface for the density matrix).  The existence of a meaningful $\zeta(x)$ depends upon this decay. For a crude approximation, one could ignore off-diagonal terms in which case $\Gamma(x,x) \approx \zeta(x)$ \footnote{For completeness, we note that the choice of $\zeta$ is not unique, but our experience on a variety of systems so far shows that the qualitative structure of $\zeta$ is quite similar even for the very coarse diagonal approximation $\zeta(x) \approx \Gamma(x,x)$ and the more intricate form of Eq. \ref{zeta_x}}.
 
Since $\Gamma$ is Hermitian, it is of interest to consider its spectral properties. Thus, we write $\hat\Gamma|\chi_\mu \rangle = \Lambda_\mu | \chi_\mu \rangle$, where $\Lambda_\mu$ and $|\chi_\mu \rangle$ are eigenvalues and eigenvectors of the operator $\Gamma(x,x^\prime)$, $\mu$ is the index which runs from 1 to the total number of grid points in three dimensional space. Then, diagonalization provides the spectral representation: $\hat{\Gamma} = \sum_\mu  | \chi_\mu \rangle\Lambda_\mu \langle \chi_\mu |$, from which we can express the conductivity as

\begin{small}
\begin{equation}
\sigma= \sum_\mu \Lambda_\mu +\sum_{x,x^\prime,x \ne x^\prime} \sum_\mu \Lambda_\mu \chi_\mu(x) \chi_\mu^*(x^\prime)
\label{spectral_sigma}
\end{equation}
\end{small}

Equation (\ref{spectral_sigma}) introduces the concept of conduction eigenvalues and conduction modes. Because of the trace invariance of $\Gamma$, the first term of Eq. (\ref{spectral_sigma}) reproduces  $\sum_x\Gamma(x,x)$.\\
\\
To implement the aforementioned method, we performed static calculations for the relaxed models with VASP. The supercell was then divided into 36$\times$36$\times$36 grid points and the wavefunction at each real space grid point was obtained from a convenient code of Feenstra and Widom \cite{widom}. We evaluated numerical gradients of wavefunctions using central difference \cite{press1992numerical} to compute $\xi_{ij}^\alpha$ for each $\alpha$. An electronic temperature of 1000 K was used for the Fermi-Dirac distribution. The $\delta$-function was approximated by a Gaussian of width 0.05 eV.

In this paper, we report the SPC for the copper-oxide models.  We are currently computing SPC calculations for the chalcogenides, and will report these elsewhere. We also know from previous experience with such materials that there is a large electron-phonon coupling for states near the gap \cite{electron_phonon_coupling} and this raises the interesting possibility of carrying out these computations at finite temperature. It is to be expected that thermal fluctuations in the structure will have an impact on conductivity and conduction pathways, as we will discuss elsewhere \cite{cu_al2o3_unpublished}.
\section{\label{sec:level3} Results and Discussions}
\subsection{\label{sec:level3_1}Conduction Mechanisms in GeSe$_3$Ag glasses}
In a recent report, we performed a direct simulation of insulator-metal transition in GeSe$_3$Ag glasses through the use of novel method of gap sculpting \cite{prasai2015sculpting} and showed that the transition to metallic phase is characterized by an apparent ``redistribution" of Se atoms from Ge(Se$_{1/2}$)$_4$ phase to Ag$_2$Se phase \cite{prasai_gese3}. In this section, we further discuss the electronic conduction mechanism and role of Ag atoms from the view point of some numerical experiments on GeSe$_3$:Ag model system. 

\subsubsection{\label{subsection_level3_1_1} A crude estimate to conduction paths in GeSe$_3$Ag glass:}

We have shown an approach to compute a real space decomposition of the conductivity in a recent work \cite{rrl_prasai18}.
In this subsection, we take a much more empirical approach and use the band-decomposed change density $q_i$($\Vec r$) 
($=|\psi_i(\Vec r)|^2$, where $i$ is the band index number) to estimate possible conduction pathways. We take the product $q_i$($\Vec r$) $\times$ $q_j$($\Vec r$) to estimate the overlap of two wave functions at $\Vec r$. The product is then modulated by electron distribution functions and the energy delta function to eliminate the overlaps that do not contribute to the conduction at a given frequency. Taking only the $\Gamma$ point ($k$=0), we compute:
\begin{equation}\label{eq_SKGF}
\begin{aligned}
{\xi}_{\omega}(\Vec r) =  \frac{ 2 \pi e^{2} \hslash^{2}}{m^{2} \omega \Omega} \sum \limits_{j=1}^{N} \sum \limits_{i=1}^{N} [f(\epsilon_{i})-f(\epsilon_{j})] \\ 
q_{i}(\Vec r) q_{j}(\Vec r) \delta(\epsilon_{j}-\epsilon_{i}-\hslash \omega)
\end{aligned}
\end{equation}
where the notations are similar to those used in Eq. (\ref{kgf_eqn}). 

\begin{figure}[!h]
{\includegraphics[width=1.65in]{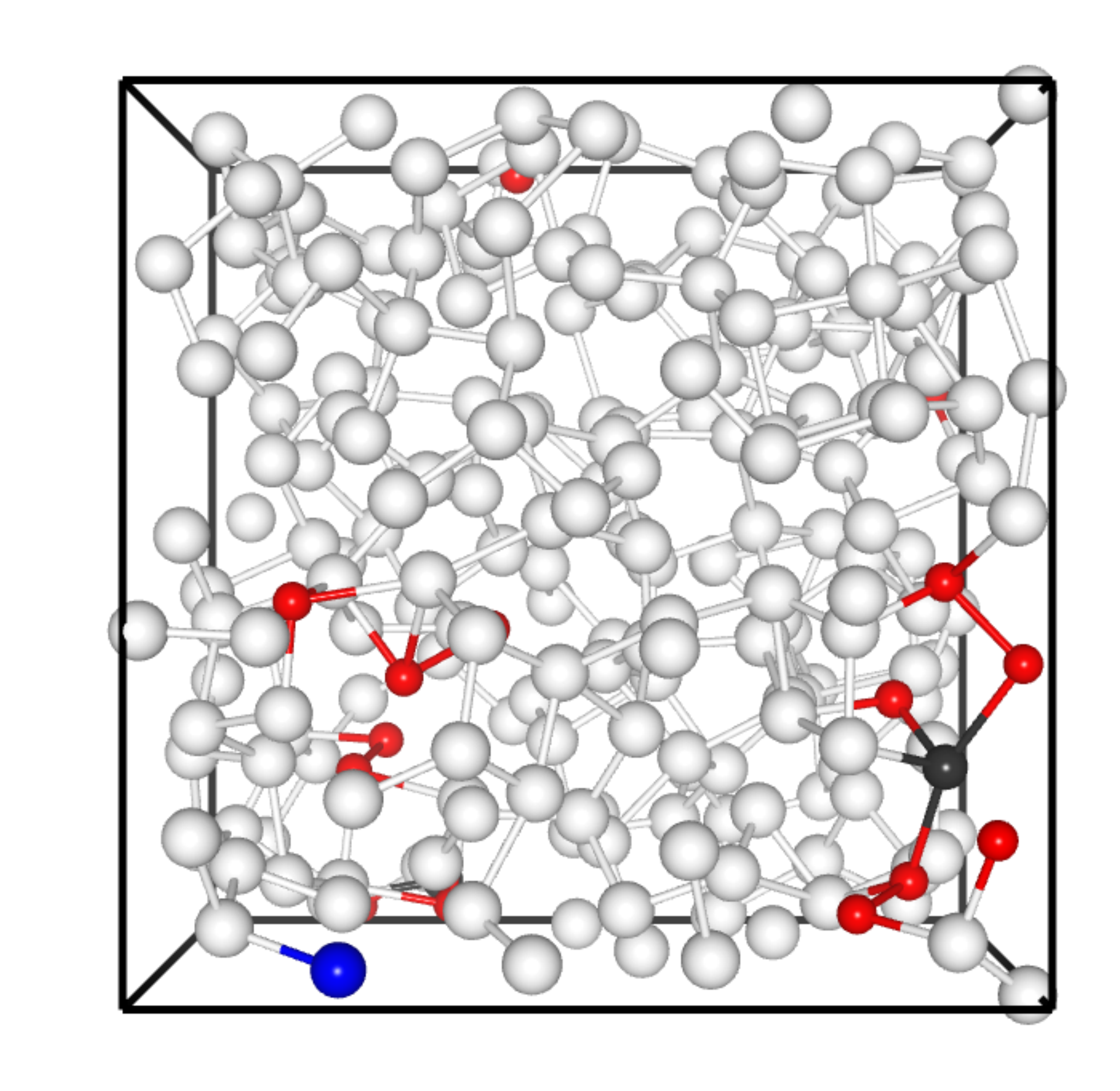}}
{\includegraphics[width=1.65in]{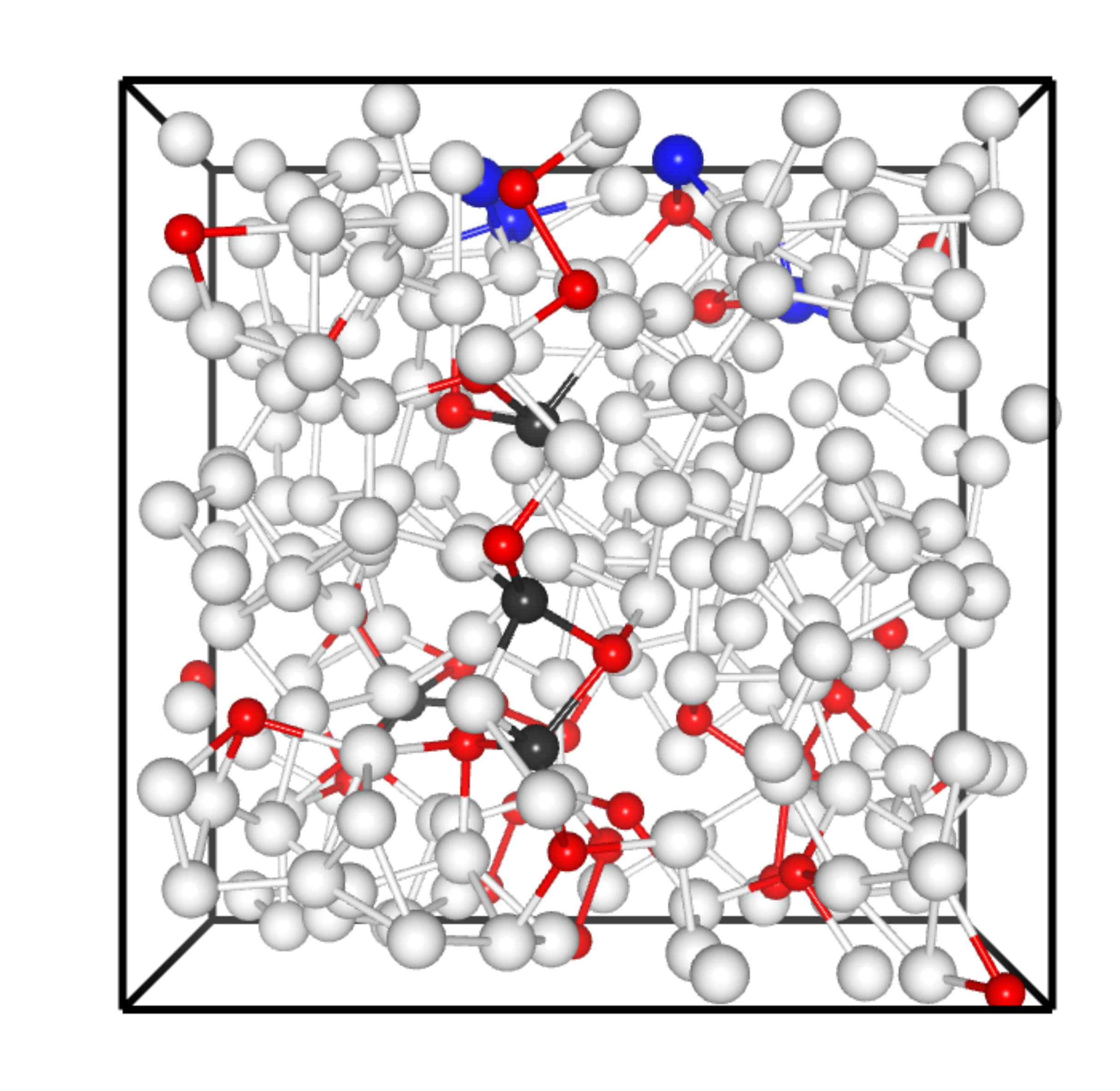}}
{\includegraphics[width=1.65in]{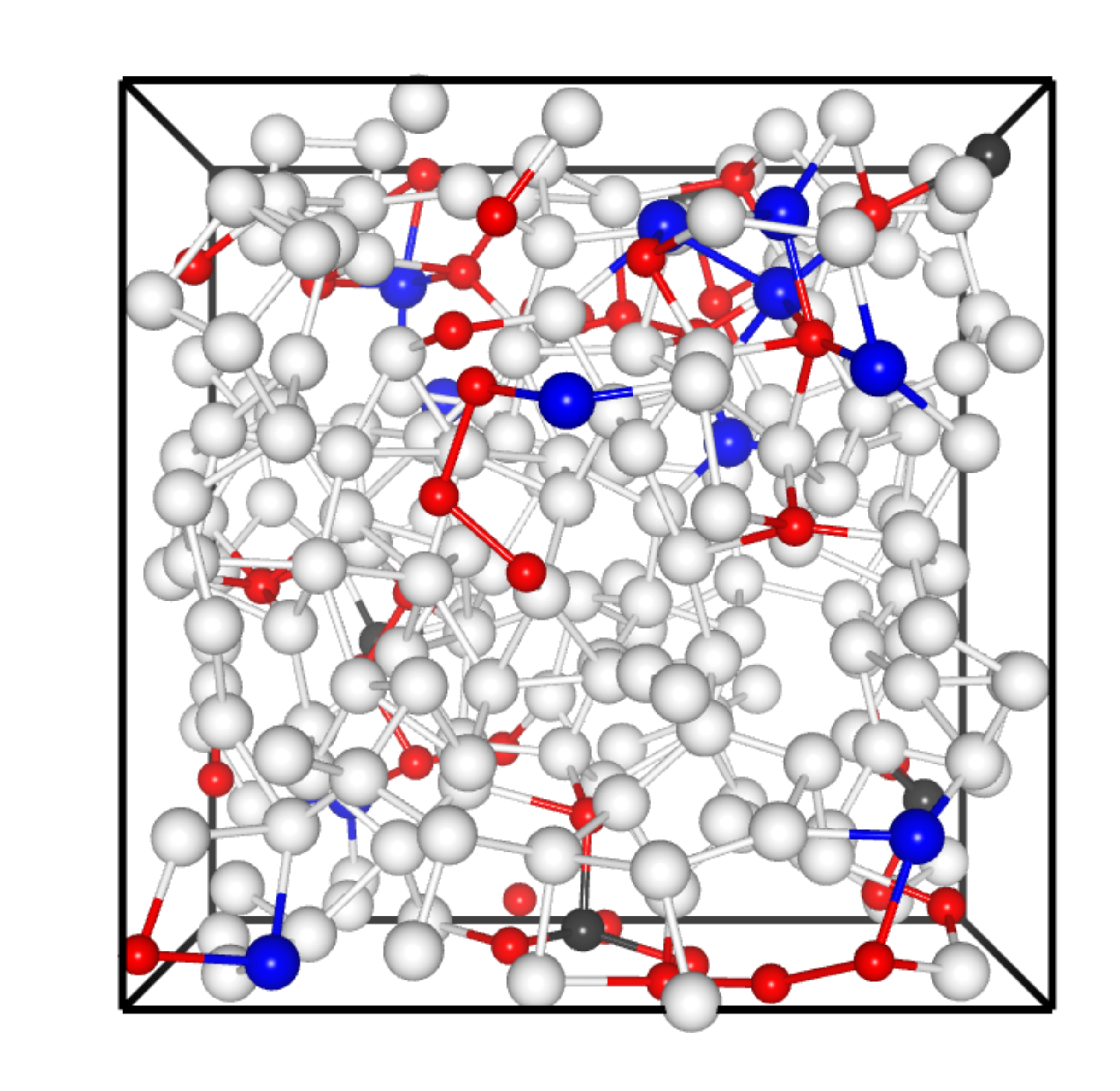}}
{\includegraphics[width=1.65in]{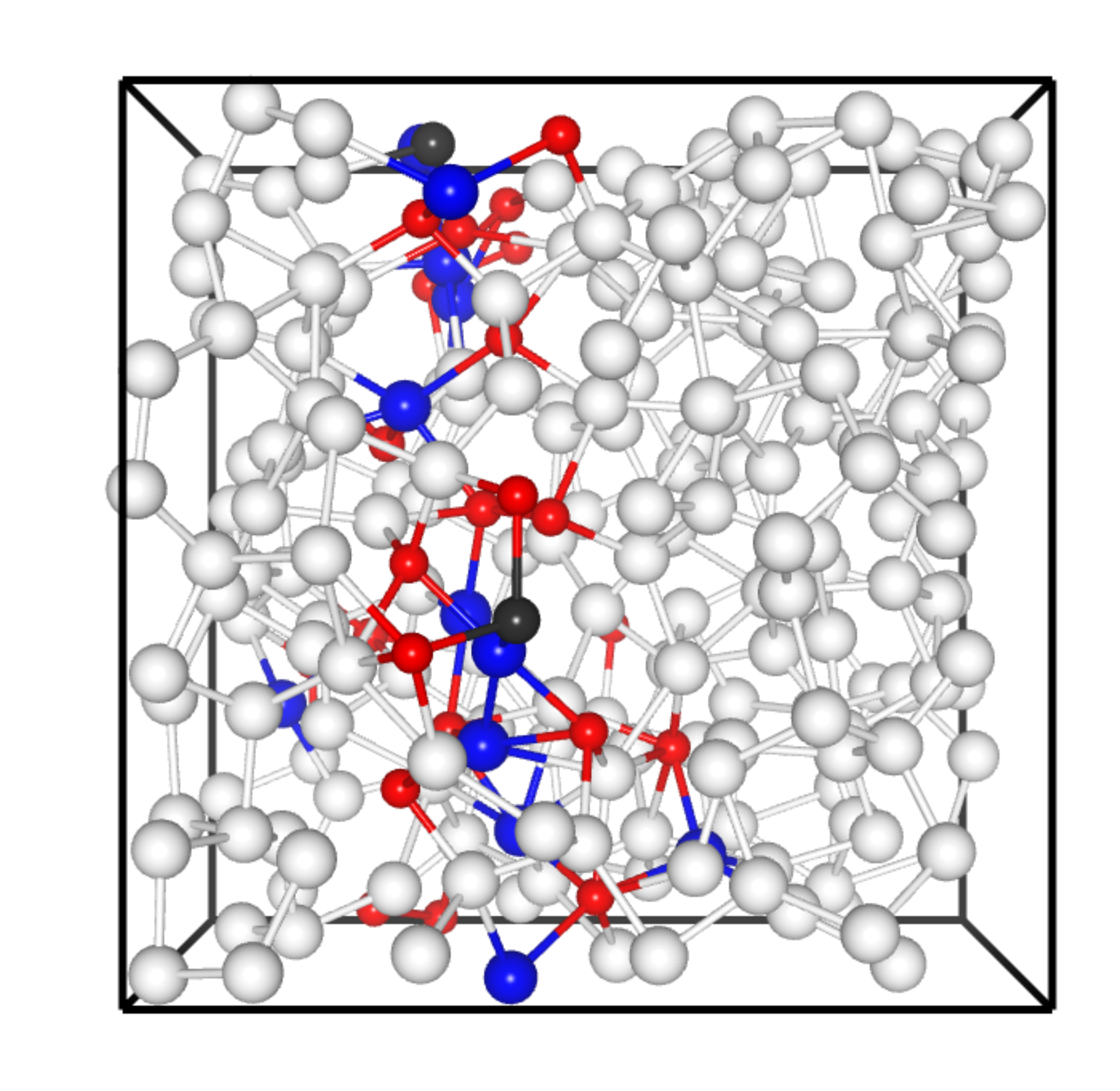}}
\caption{The possible conducting parts of GeSe$_3$Ag models for Ag concentrations at the DC limit. From left, TOP: 10\% and 15\% and BOTTOM: 25\% and 35\%. Equation (\ref{eq_SKGF}) is used to compute space projected conductivity which is then projected on the atomic sites. Color nomenclature: Red Se atoms, Blue Ag atoms and Black Ge atoms.} 
\label{spc_bader_geseag} 
\end{figure}

For each $\omega$, the above equation crudely highlights the parts of the network that are likely to conduct. The values of ${\xi}_{\omega}(\Vec r)$ can then be projected onto the atomic sites and the relative values of projections would indicate which atoms contribute directly to electron transport at a given frequency. Figure \ref{spc_bader_geseag} shows a realization of such calculation where the atoms that have highest projection of ${\xi}_{\omega}(\Vec r)$ (i.e. are likely to directly contribute in electron transport) are highlighted in color. We observe that Se-atoms play the most significant role in conduction for lower Ag-concentrations. At higher Ag-concentration, Ag atoms begin to play more roles but Se atoms still remain dominant.

\subsubsection{\label{subsection_level3_1_2} Ag-nanowires in glassy matrix}
\begin{figure}[!h]
\begin{center}
\includegraphics[width=1.65in]{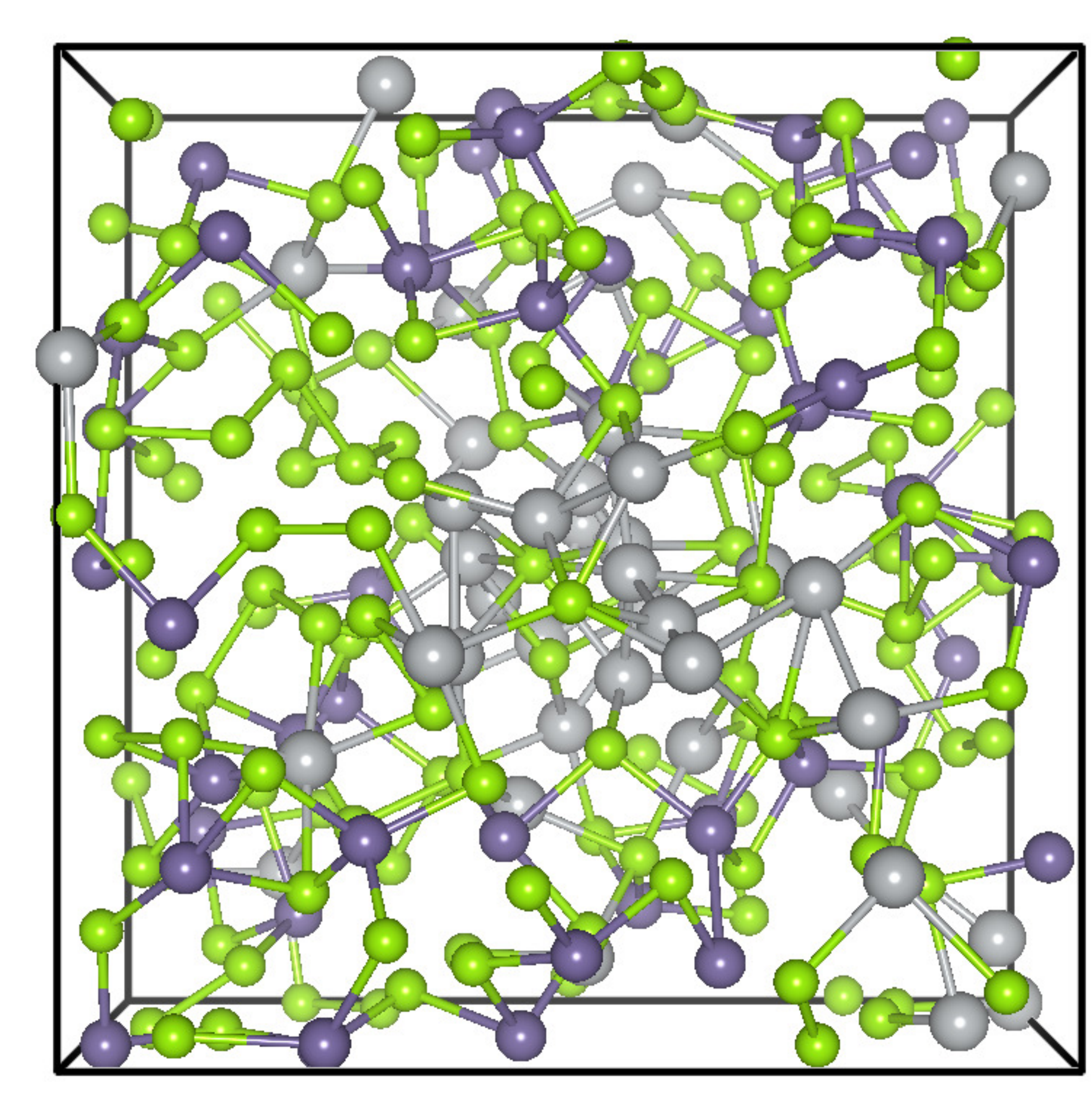}
\includegraphics[width=1.65in]{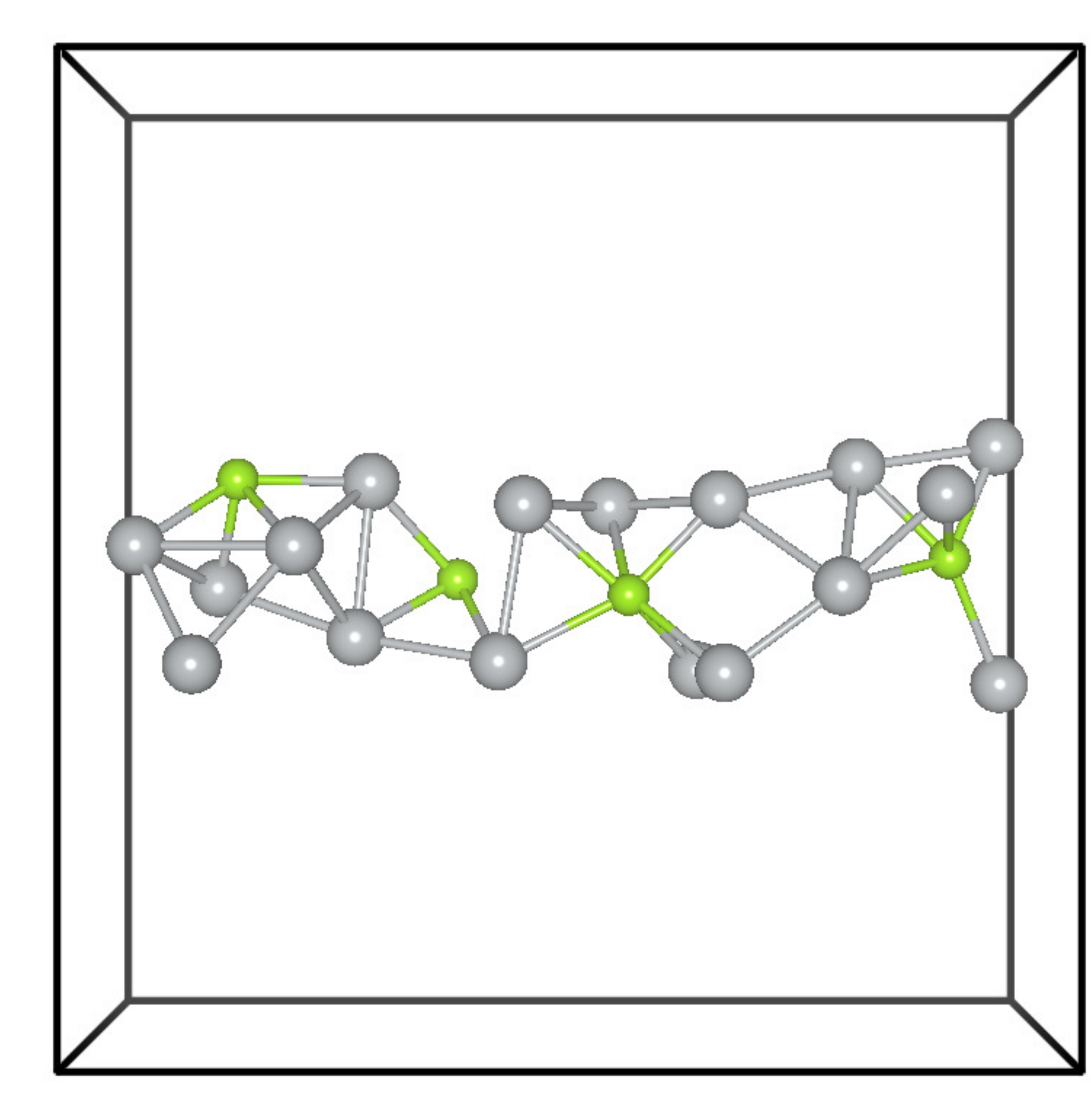}
\includegraphics[width=1.65in]{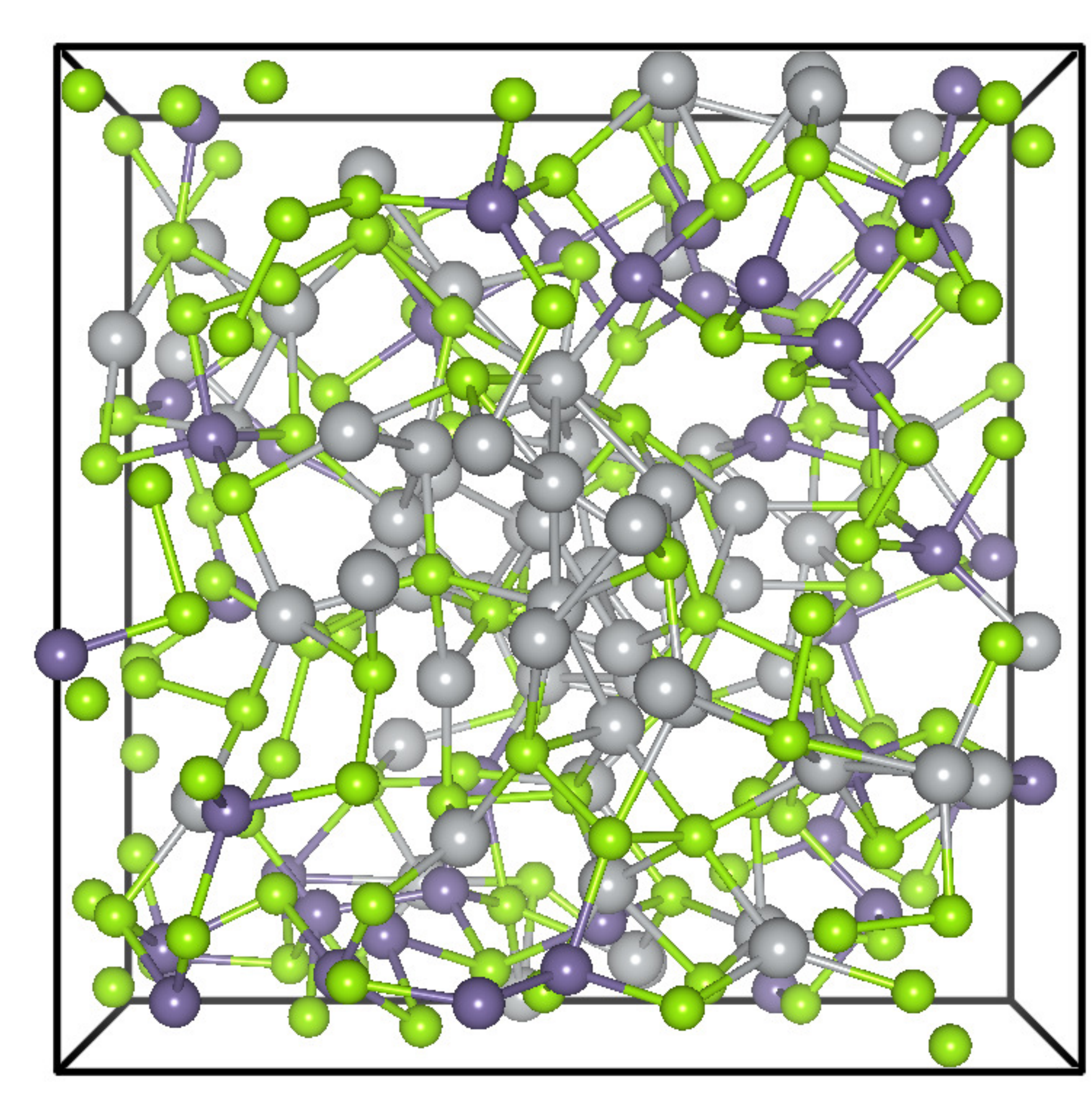}
\includegraphics[width=1.65in]{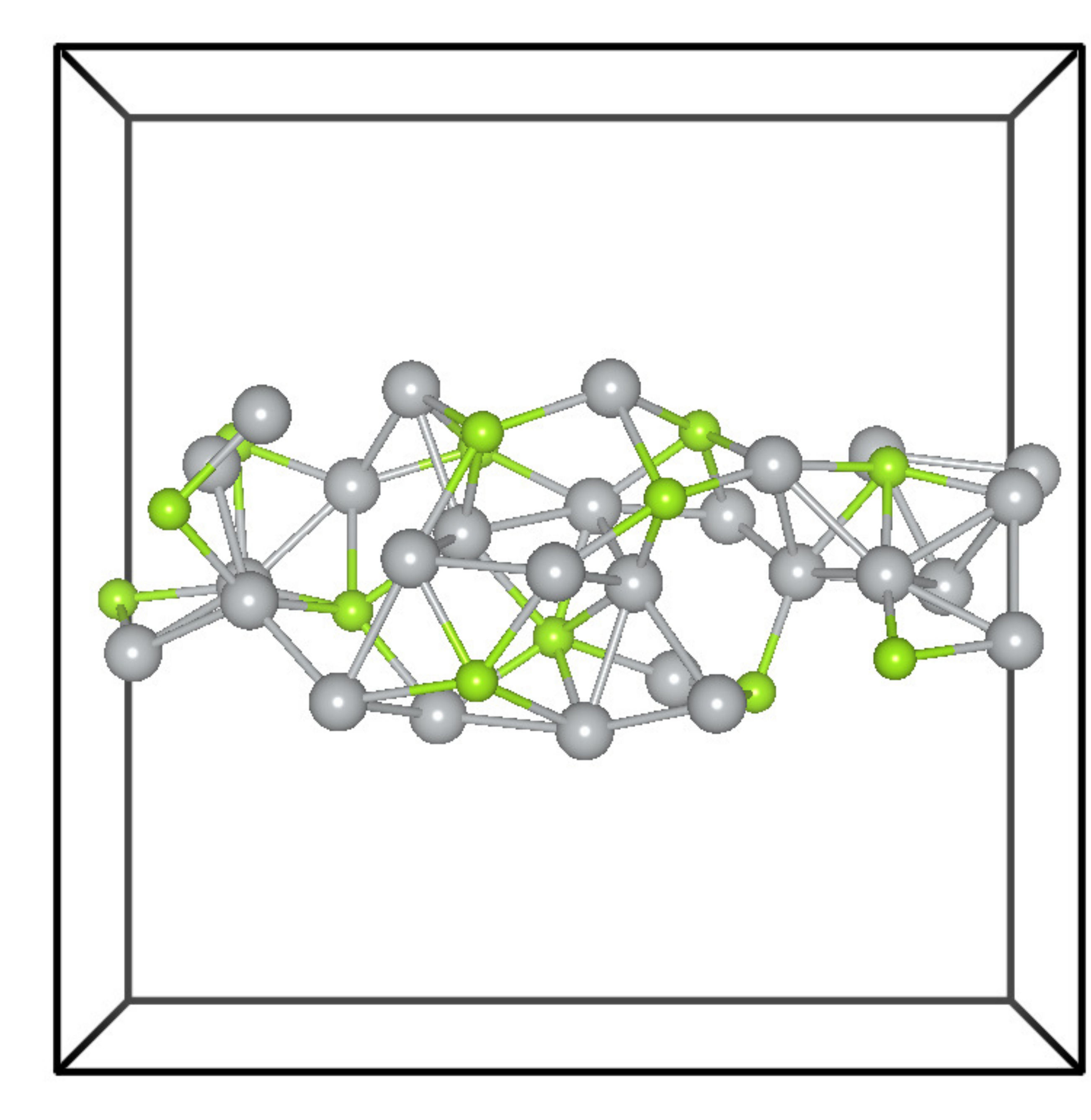}
  \caption{ (Top left) Model A with Ag-filament viewed parallel to the axis of the filament.
      (Top right) The central Ag filament in the model A. 
      (Bottom left) Model B with Ag-filament viewed parallel to the axis of the filament.
      (Bottom right) The central Ag filament in the model B. Color nomenclature: Ge:Purple, Se: Green, Ag: Silver }
  \label{ag_nanowire}
  \end{center}
\end{figure}

In this section, we discuss the role of Ag-nanowires on electronic transport based on the results of explicit simulations of the Ag-filaments in GeSe$_3$Ag glass matrix. We employ a model with 10\% Ag (among the ones discussed in section \ref{sec:level2_1}) and use constrained AIMD to install a Ag-nanowire at the center of the model. A rectangular block that connects two opposite faces of the cubic model is taken and the atoms within the block were identified as Ag. The block has dimensions of 5.06 \AA~$\times$ 5.06 \AA~$\times$ 18.60 \AA. The cross-section of this block is chosen in order to preserve the overall n$_{Ge}$ to n$_{Se}$ ratio. The final stoichiometry of the wire-matrix system is (GeSe$_3$)$_{0.84}$Ag$_{0.16}$. This model is then annealed at 700 K for 5 ps, then at 1000 K for 5 ps and finally quenched to 300 K in 4.5 ps.  During the annealing and quenching cycle, the Ag atoms in the rectangular block are constrained to move only along the axis of block. The quenched structure is then relaxed to its nearest energy minimum using {\it unconstrained} conjugate gradient moves. The energy minimized model has a continuous Ag filament across the center of the cell (figure \ref{ag_nanowire}) and the glassy backbone bonds well with the filament so that no internal surfaces are formed in the boundary between the filament and the matrix (figure \ref{ag_nanowire}). The volume is kept fixed since the density difference between the systems with $x$=10 and $x$=15 is small. The pressure in the model is found to be 0.02 GPa which is effectively zero pressure. For the sake of convenience, we label this model as `Model A'.

We then repeat the same process starting with a model with 15\% Ag discussed in section \ref{sec:level2_1} and install a nanowire in rectangular block of dimension 7.00 \AA $\times$ 7.00 \AA $\times$ 18.65 \AA. The final stoichiometry of the wire-matrix system for this system is (GeSe$_3$)$_{0.74}$Ag$_{0.26}$. In the discussion that follows, we call this model `Model B'. 

\begin{figure}[ht]
\begin{center}
{\includegraphics[width=3in]{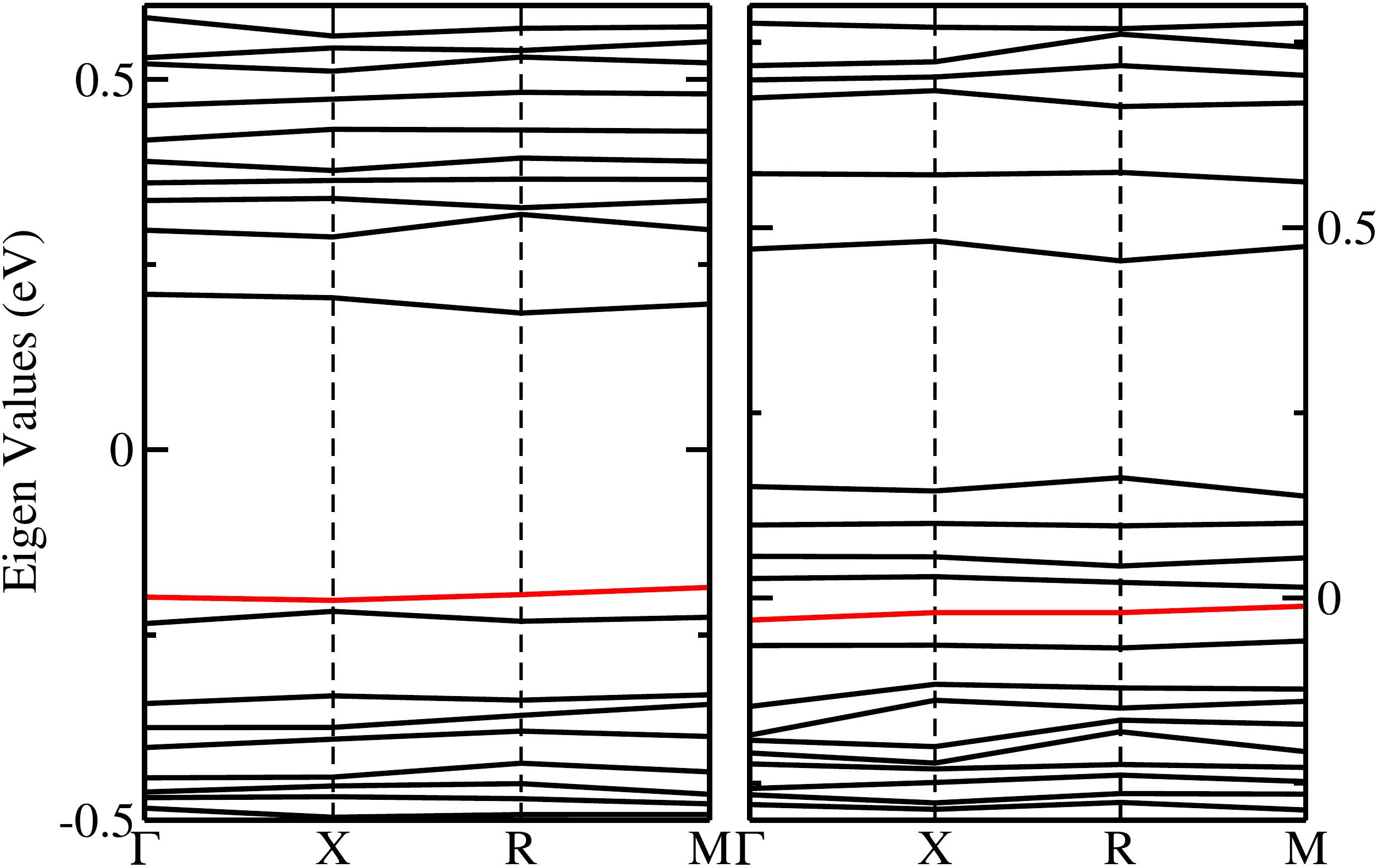}}
\caption{The band structure of model A (right) compared with homogeneous model with close stoichiometry (left). The band structure on the left is of a homogeneous model with composition (GeSe$_3$)$_{0.85}$Ag$_{0.15}$ and that on the right is of the model A (i.e the model with Ag nanowire and with composition  (GeSe$_3$)$_{0.84}$Ag$_{0.16}$). The Fermi energy is at zero and HOMO level is highlighted by red line.} 
\label{fig4} 
\end{center}
\end{figure}

\begin{figure}[ht]
\begin{center}
{\includegraphics[width=3in]{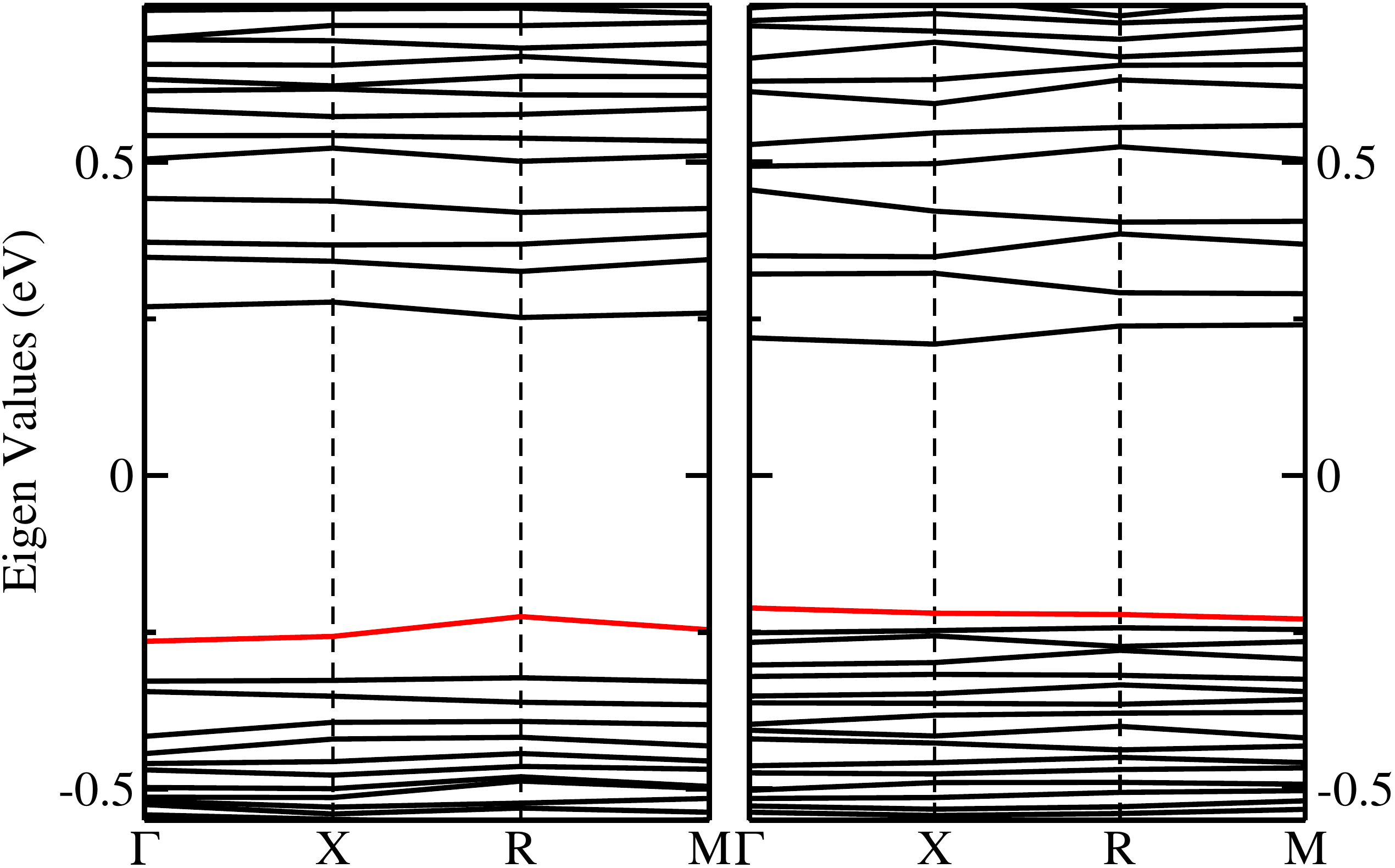}}
\caption{The band structure of model B (right) compared with homogeneous model with close stoichiometry (left). The band structure on the left is of a homogeneous model with composition (GeSe$_3$)$_{0.75}$Ag$_{0.25}$ and that on the right is of the model A (i.e the model with Ag nanowire and with composition  (GeSe$_3$)$_{0.74}$Ag$_{0.26}$). The Fermi energy is at zero and HOMO level is highlighted by red line.} 
\label{fig5} 
\end{center}
\end{figure}

\begin{figure}[ht]
\begin{center}
{\includegraphics[width=3in]{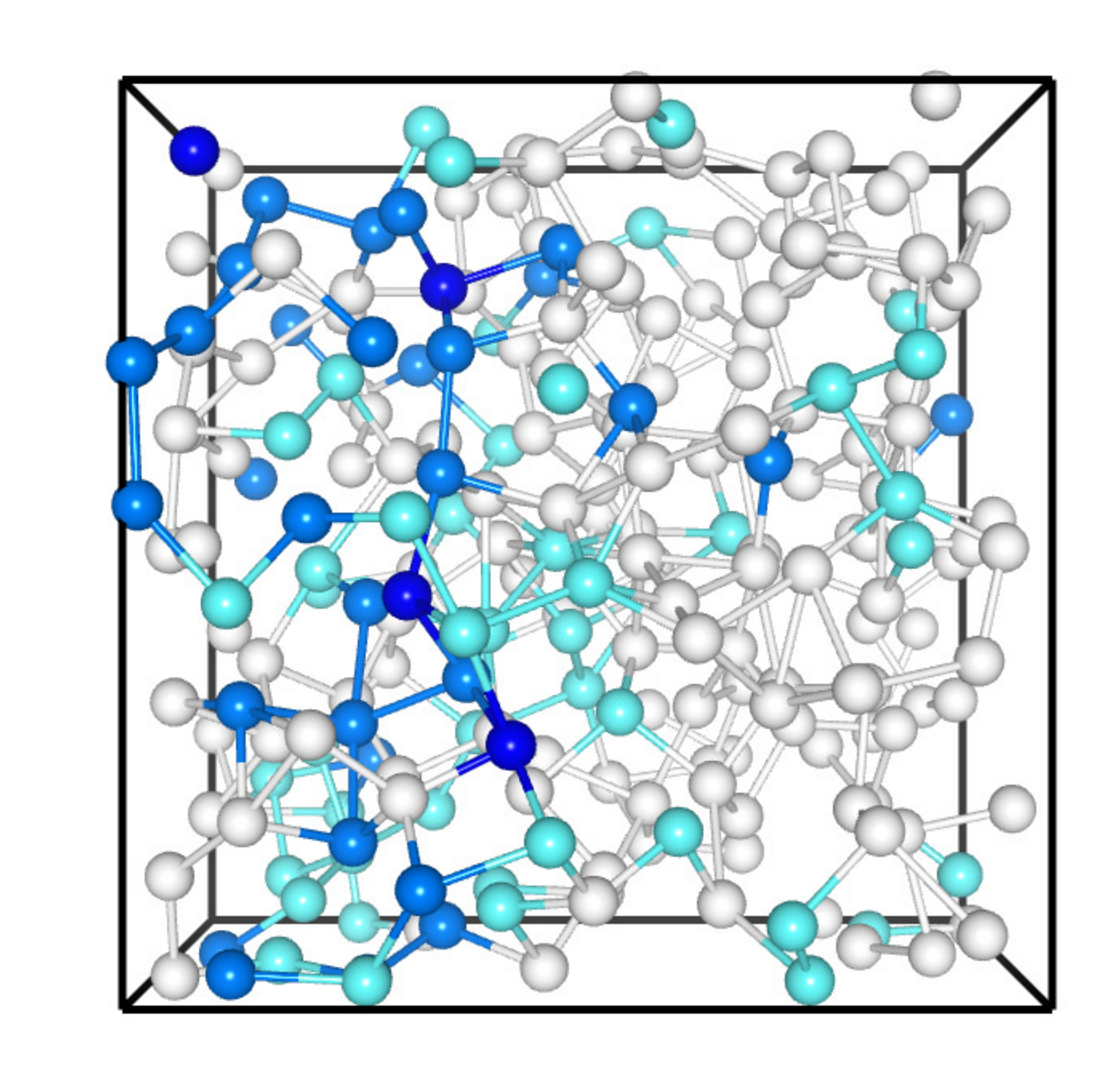}}
\caption{The resonant clusters in model A. The top 40 atoms (out of 240) contributing to
          three doping levels with energy $\epsilon_F$+0.055 eV, $\epsilon_F$+0.096 eV and $\epsilon_F$+0.150 eV
          are presented in color. Those atoms that fall in top 40 of {\it all} three states (two of three
          states/one of three states) are painted in dark blue (light blue/ice blue).} 
\label{fig15} % common label
\end{center}
\end{figure}

We compute the electronic structure of the systems by sampling over 4 points on the Brillouin zone \cite{monkhorst_pack}. As expected, there is negligible k-dispersion since the size of the unit cell is fairly large. Figure \ref{fig4} and \ref{fig5} show band structures of model A and model B alongside homogeneous models of comparable composition. Interestingly, we observe that Ag-nanowire does {\it not} produce a metal-like DOS but rather all models show a well defined band gap. We find that model A is slightly p-doped. The HOMO-LUMO gap in the model with Ag-filament
is 0.04 eV. The optical gap is 0.29 eV which is still smaller compared to 0.38 eV of model without the Ag-filament. For model B, the HOMO-LUMO gap is 0.43 eV which is smaller compared to 0.53 eV of the model without the Ag-filament.  These gaps are calculated using uncorrected DFT which tends to underestimate the optical gap. Bader charge analysis \cite{bader1998atoms} shows that Ag-atoms in the filament have similar charged state as the Ag-atoms in the glass backbone.
It is particularly interesting to note the presence of three impurity states in model-A. 
The calculated conductivity of model-A using Kubo-Greenwood formula (equation \ref{kgf_eqn}) is higher than that of the homogeneous model of similar stoichiometry by a factor of 10$^8$ (the details on our calculation of conductivity can be found in \cite{prasai_gese3}).
In order to understand the role the impurity states in model-A might play in the transport,
we investigate the structural origins of these 3 doping levels. Note that these levels are nearly degenerate with energies  $\epsilon_F$+0.055 eV, $\epsilon_F$+0.096 eV and $\epsilon_F$+0.150 eV. A projection of these states onto atomic sites shows that these levels have significant spatial overlap (see Fig. \ref{fig15}). This suggests that there is mixing between the doping levels and these levels form an extended band. Such a system should enable electronic conduction since an electron is more likely to find an overlapping state with similar energy. The likely mechanism of conduction from this line of reasoning is the hopping transport through the resonant clusters, discussed in more detail elsewhere \cite{dong1998}. 
\subsubsection{\label{subsection_level3_1_3} The role of non-bridging Se atoms}
\begin{figure}[ht]
\begin{center}
{\includegraphics[width=3in]{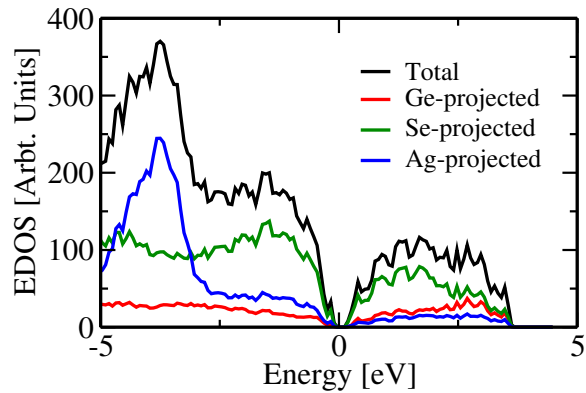}}
\caption{Projected density of states (DOS) on atomic sites for (GeSe$_3$)$_{0.85}$Ag$_{0.15}$ models The Fermi level is shifted to zero.} 
\label{pdos_geseag} % common label
\end{center}
\end{figure}

\begin{figure}[ht]
\begin{center}
{\includegraphics[width=3in]{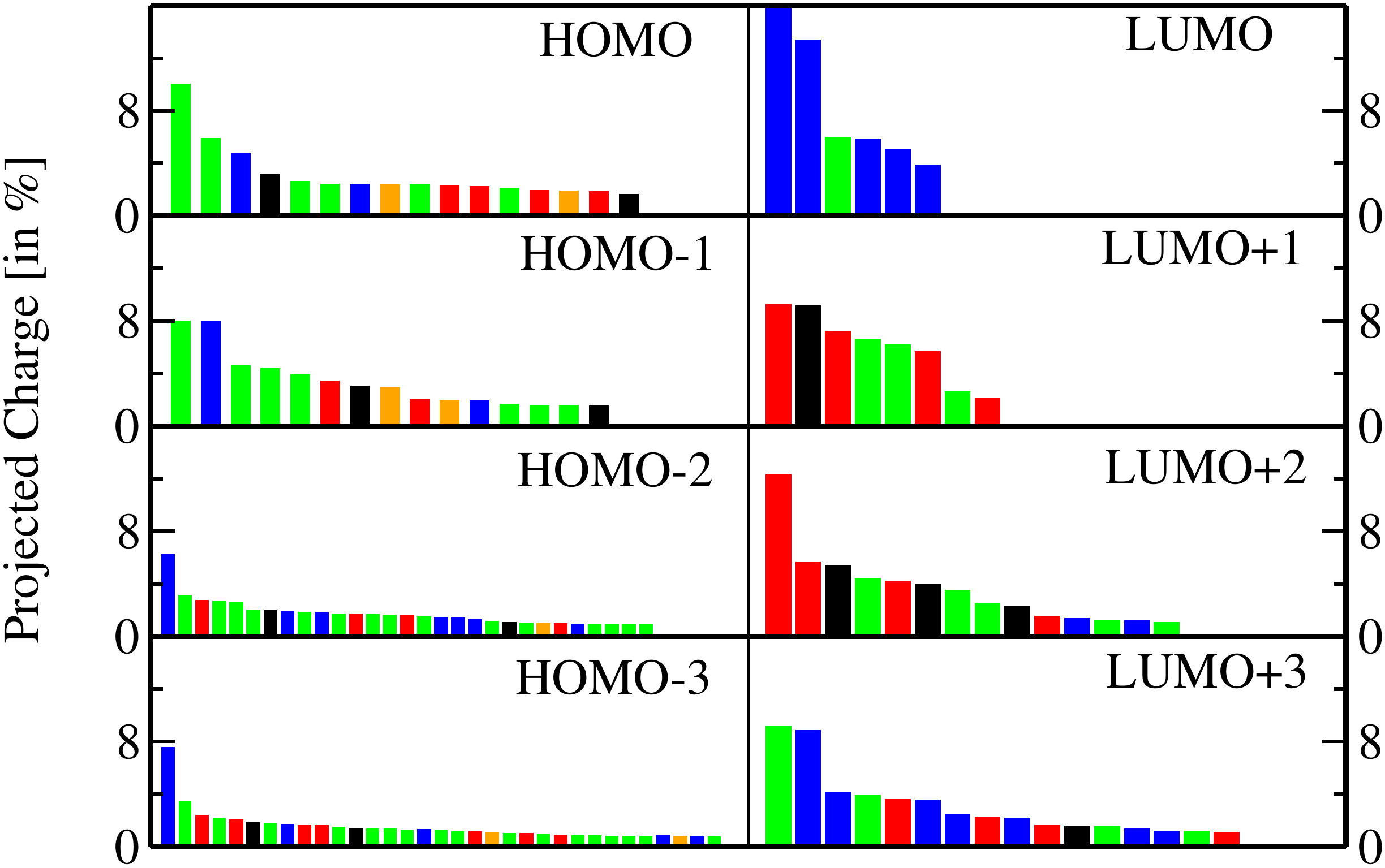}}

\caption{The projected charges of band edge states onto the atoms. Eight bands in
the band-gap region are presented here. HOMO and LUMO represent the valence edge
and conduction edge respectively. Each rectangle in the histogram represents an atomic
site where the color is used to denote atom environments: Bridging Se-atoms (Red), Non-
bridging Se-atoms (Green and Blue, Green for Se atoms bonded with one Ge-atom and
Blue for Se-atoms not bonded to any Ge atom). Black and Orange colors represent Ge and
Ag atoms respectively. Only the atoms with highest contribution to the band up to 50\% of
total charge of the band (i.e. 2$e^{-1}$) are shown.}
\label{major4} % common label
\end{center}
\end{figure}

 In light of the discussion in sections \ref{subsection_level3_1_1} and  \ref{subsection_level3_1_2}, as well as our earlier work \cite{prasai_gese3}, it is clear that Se-atoms play a significant role in transport behavior of these glasses. We observe that the states in the band edges are derived mainly from Se p-orbitals as shown from the projected density of states in Fig \ref{pdos_geseag}.
 Se atoms exhibit a wide range of local environments, bonding with Ge to form host network, bonding with Ag atoms to form a Ag$_2$Se phase, and attaching to other Se atoms to form Se-filaments. These bonding environments lead to different electronic activity around the Fermi level. To discuss this in familiar terms, we use the terminology commonly used to describe O atoms in common network-former/ network-modifier glasses (e.g. NaSiO$_3$). We categorize Se-atoms bonded with 2 Ge atoms as 'bridging' Se atoms. Similarly, Se-atoms bonded to one or no Ge-atom are referred here as 'non-bridging' Se-atoms. The later group of Se-atoms either terminates the GeSe network (one-fold Se) or is more or less mobile and form AgSe phase/short Se-chains. With this classification in mind, it is interesting to observe that the non-bridging Se atoms are highly electronically active around Fermi-energy and are transport significant. We make this point in Fig. \ref{major4} where we show the top projections of eight band-edge states on to the atoms. It is clear that the non-bridging Se-atoms play a key role band edges.
 This also aligns with the well-known observation that states deeper into the band are more extended as can be noted from the higher number of atom projections used to amass top 50\% of the charge for those states. We also note the asymmetrical localization on the two sides of Fermi level. Conduction edge states appear to be more localized than the valence edge states. The anomalous origin of LUMO+1 and LUMO+2 comes from two strained tetrahedral Ge sites.
% Here comes the Alumina stuff
\subsection{Conduction Mechanisms in Al$_2$O$_3$:Cu models}
\begin{figure}[!h]
 \begin{center}
 \includegraphics[width=1.65in]{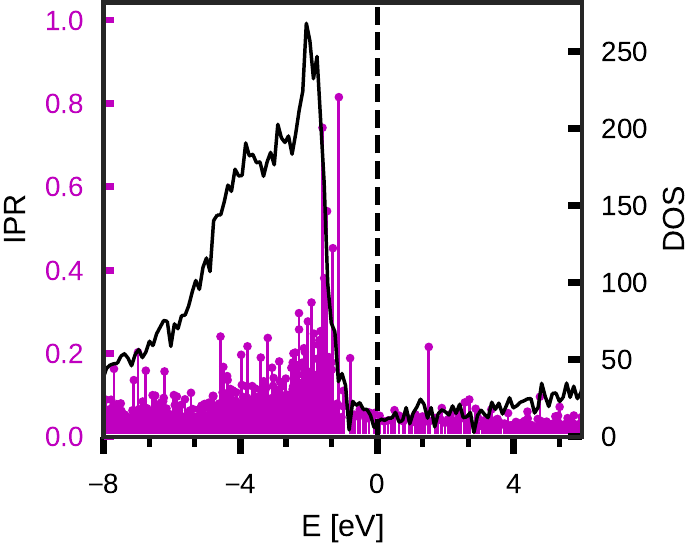}
 \includegraphics[width=1.65in]{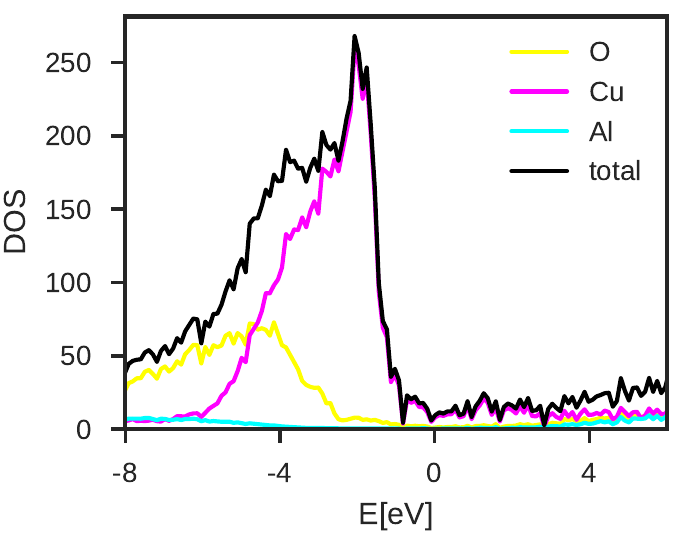}
 \caption{ LEFT: Total density of states and the inverse participation ratio (IPR). RIGHT: Density of states projected on atomic sites for 30\% Cu. The Fermi level is shifted to zero in both figures.}
 \label{ipr_dos_pdos_cu_30_fig}
 \end{center}
 \end{figure}

\begin{figure}[!h]
 \begin{center}
 \includegraphics[width=1.65in]{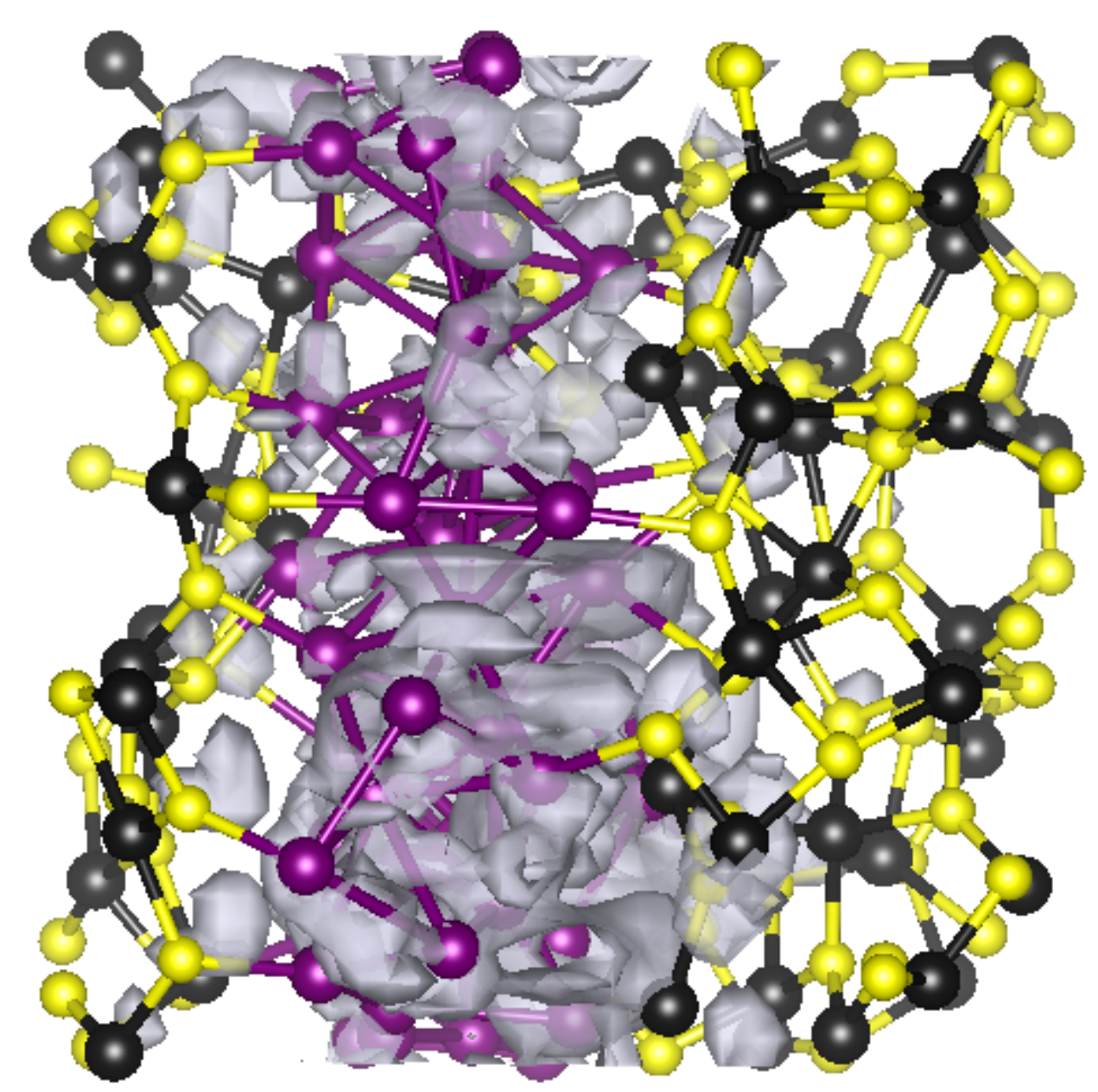}
 \includegraphics[width=1.65in]{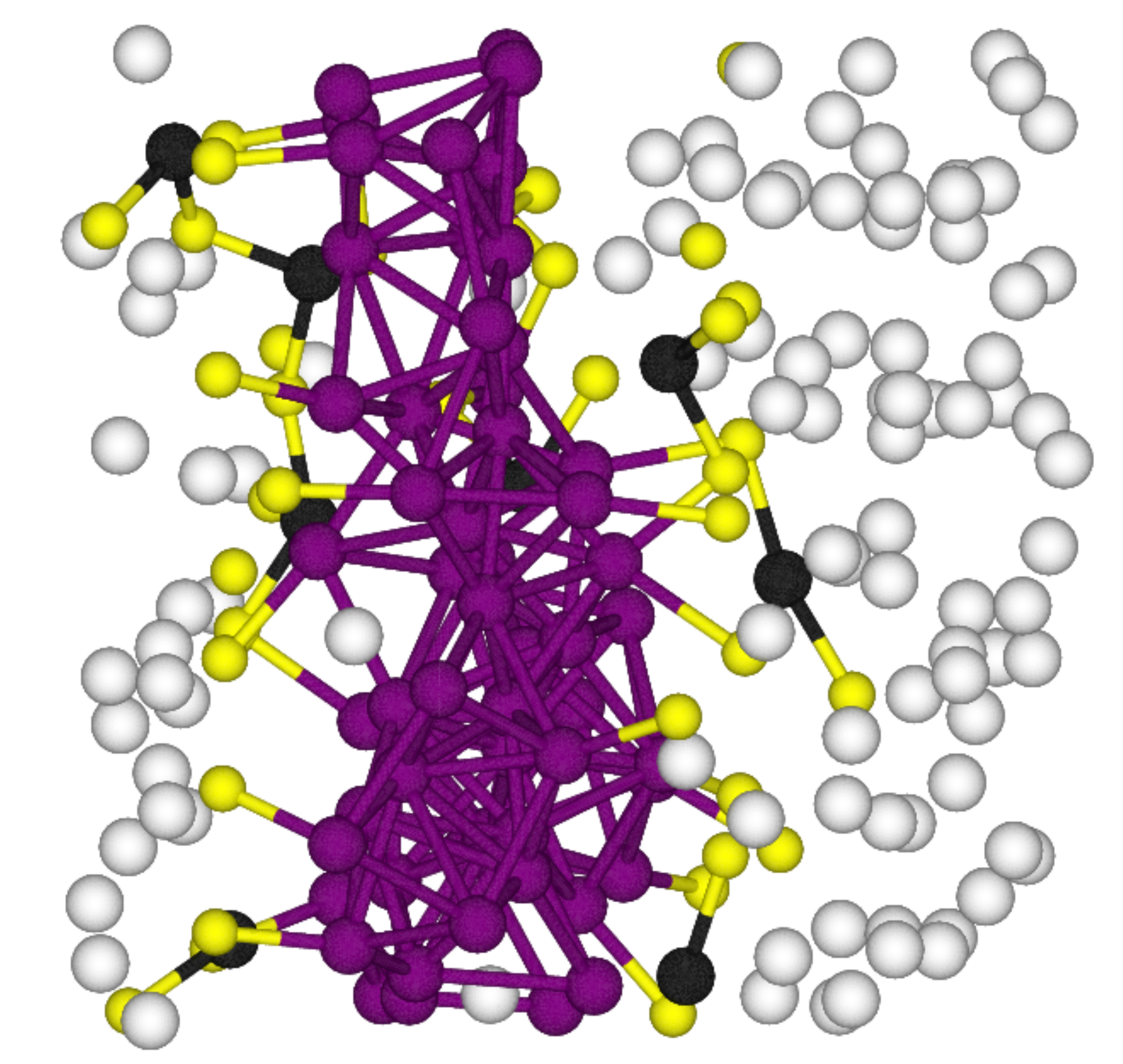}
 \caption{ LEFT: Isosurface plot of SPC (white blob) obtained from $\zeta(x)$. Atom color Magenta: Cu, Yellow: O and Black: Al. RIGHT: Bader projection of conductivity projected on atomic sites. The cutoff of 98\% is chosen. Same color code as left plot. The white colored atoms with values less than the cutoff.}
 \label{spc_bader_gamma_30cu}
 \end{center}
 \end{figure}

\begin{figure}[!ht]
\begin{center}
{\includegraphics[width=3in]{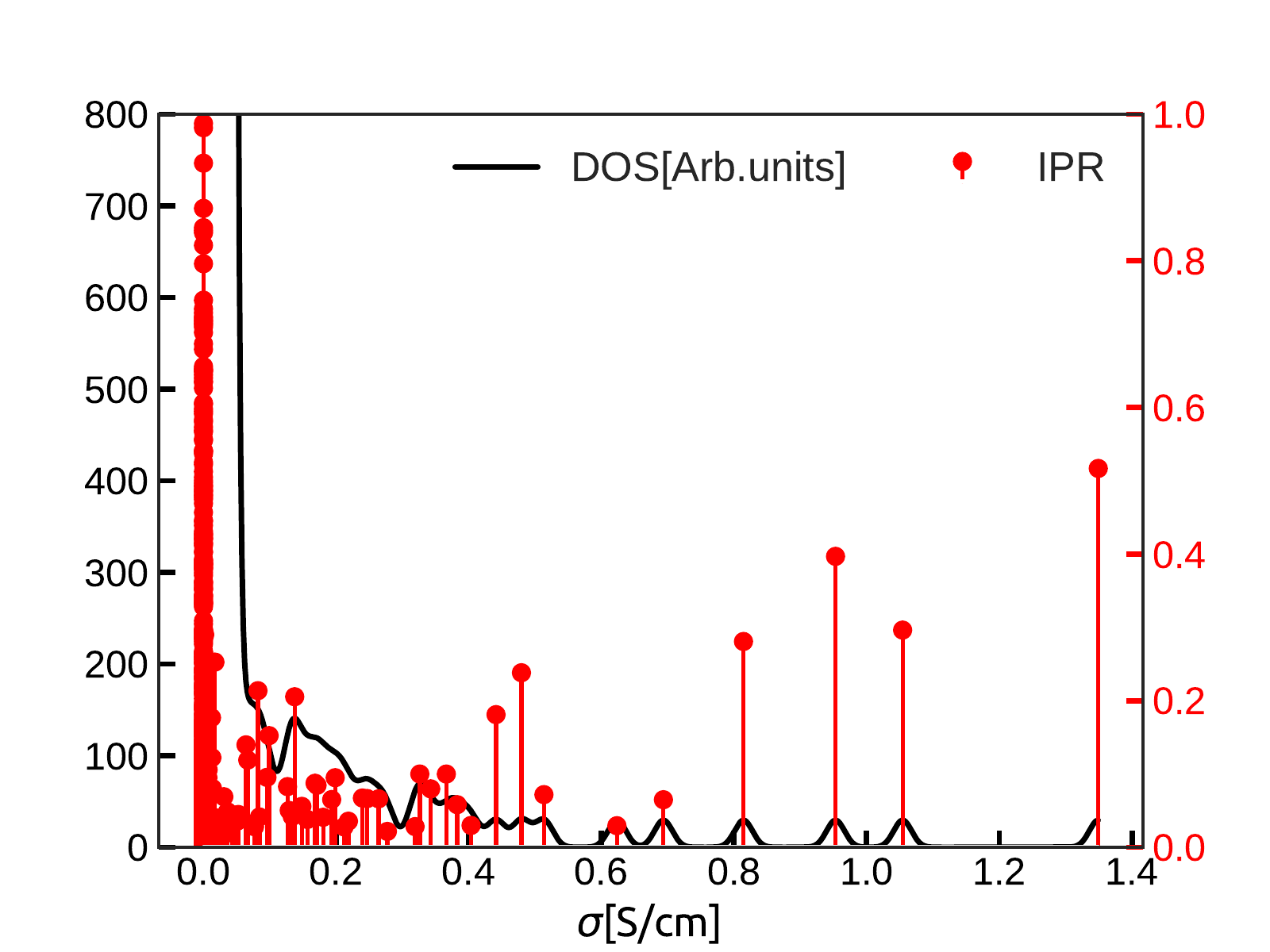}}

\caption{ Spectrum of $\Gamma$ and localization (IPR) of its eigenvectors ($\chi$) for 30\% Cu model.} 
\label{ipr_dos_gamma_30cu} % common label
\end{center}
\end{figure}

 The story of Cu in alumina is different than that of Ag in GeSe$_3$.  The Cu atoms segregate and form clusters in alumina, whereas Ag atoms organize themselves more or less uniformly in GeSe$_3$. The highly ionic character of the alumina host (contrasted with the relatively covalent selenide) is the origin of the difference. Indeed, the tendency to cluster is associated with the host rather than the transition metal, since we have shown separately that Ag clusters in the alumina host much like Cu does.  As Cu is introduced into the alumina host, levels appear in the alumina gap. Quite interestingly, the gap fills up rather {\it uniformly} as Cu concentration increases, one does not see an ``impurity band" form, as one might suppose from experience with doped semiconductors. Because the Cu clusters, Cu regions eventually become space filling (with periodic boundary conditions) and these connected filaments become pathways for metallic conduction. The models with 20\% and 30\% Cu are conducting. Here, we focus our study on the 30\% Cu model. Left plot in Fig. \ref{ipr_dos_pdos_cu_30_fig} shows the total density of states and IPR. The states near the Fermi level are fairly extended, and there is a continuous filling of states in the fundamental electronic (optical)  gap of the oxide host. The right plot in Fig. \ref{ipr_dos_pdos_cu_30_fig} shows the density of states projected on atomic sites for this model which reveals that the states near Fermi level are mostly derived from Cu atoms. This case is quite different from the Ag in GeSe$_3$ systems, where most of the contribution to the total density of states come from Se atoms as shown in Fig. \ref{pdos_geseag}.\\
The usual means to determine the conducting parts of a network is to just look at the Kohn-Sham orbitals near the Fermi level in the range of order $k$T, where $k$ is the Boltzmann constant, as we did in figure \ref{spc_bader_geseag}. This is at best qualitative, and the present work correctly includes effects stemming from the current-current correlation function, which underlies Kubo's approach to electrical transport. The left plot in Fig. \ref{spc_bader_gamma_30cu} visualizes an iso-surface plot of space projected conductivity (SPC) using Eq. \ref{zeta_x} showing the transport-active components of the network for Cu-concentration 30\%. We see that the electronic conduction is primarily along connected Cu atoms. Few O atoms which are basically bonded to the Cu atoms also participate in conduction process whereas Al atoms do not show their active contribution in this process. \\
The diagonalization of $\Gamma$ yields the conduction eigenvalues and modes. The right plot in Fig. \ref{spc_bader_gamma_30cu} shows the Bader projection of the weighted conductivity on the atomic sites obtained from Eq. (\ref{spectral_sigma}). We see that the conduction is through the Cu chain of the network. Thus, connectivity of Cu-Cu dramatically influences the electronic conduction in these materials. Figure \ref{ipr_dos_gamma_30cu} reports the density of states for $\Gamma$ and spatial localization of of the eigenvectors $\chi_\Lambda(x)$. The eigenvectors conjugate to extremal (large) $\Lambda$ are highly extended. Few eigenvalues contribute significantly to $\sigma$, with an overwhelming accumulation of spectral weight in the density of states near $\Lambda=0$. We observe spectral tail in the density of states for this system near $\Lambda$ = 0. We find it interesting that the eigenvectors of $\Gamma$ naturally categorize the total conductivity into a tiny (compared to the dimension of $\Gamma$) collection of spatially resolved contributions \footnote{ In other calculations not reported here, we have found that qualitatively, the density of states for $\Gamma$ always has huge spectral weight at $\Lambda=0$, and for insulating or weakly conducting systems, a handful of non-zero $\Lambda$ appear with small IPR (meaning that they are highly extended) and the spectral accumulation near $\Lambda=0$ involves only highly localized states. Thus, diagonalizing $\Gamma$ leads naturally to the dominant few modes of conduction, and we expect this to be useful in many systems in the future. It is interesting that for a true metallic systems such as crystalline Al, a spectral tail forms. The extended state conduction through the Cu in the 30\% alumina model also exhibits this tendency.}.\\

\section{Conclusion}
In this paper we determined the atomistic origin of electrical conduction in  chalcogenide and oxide CBRAM materials. The strong tendency of Cu or Ag to segregate in the oxide host (the opposite of what happens in chalcogenide hosts) has important implications to transport and cell operation. It is intriguing that the addition of Cu to the oxide leads to a uniformly filled host optical gap and ultimately metallic conduction. To oversimplify, it appears from this work that for CBRAM applications, one has to ``work hard" (electrochemically) to draw Ag together in a chalcogenide host, whereas in the oxide the labour is in breaking apart Cu clusters. Of course it would be very helpful to carry out electrochemical simulations to explore this further, though this might require techniques less CPU intensive than VASP. We also observe that a conventional metallic form of conduction seems to explain transport in the copper oxide, whereas something more akin to hopping seems to be operative for Ag. Very thin nanowires of Ag do {\it not} metallically conduct in the chalcogenide. Of course for broad enough Ag filaments conduction must occur, and this appears to be another stark contrast with the alumina system. Indeed, our best success at obtaining an {\it ab initio} model of a conducting silver chalcogenide system used our \enquote{gap sculpting} method, as we discuss elsewhere \cite{prasai2015sculpting}.
\section{Acknowledgments}
 We thank the US NSF for support under grants DMR 1507670 and 1506836. Also, some of this work used the Extreme Science and Engineering Discovery Environment (XSEDE), which is supported by National Science Foundation grant number ACI-1548562, using BRIDGES at the Pittsburgh Supercomputer Center under the allocation TG-DMR180083. We have benefited from many conversations with Profs. Michael Kozicki, Maria Mitkova and Gang Chen.\\
\input{main.bbl}
%\bibliography{reference}
\end{document}

%% file: main.bbl
%merlin.mbs apsrev4-1.bst 2010-07-25 4.21a (PWD, AO, DPC) hacked
%Control: key (0)
%Control: author (8) initials jnrlst
%Control: editor formatted (1) identically to author
%Control: production of article title (-1) disabled
%Control: page (0) single
%Control: year (1) truncated
%Control: production of eprint (0) enabled
%